\documentclass[twocolumn,english,prl]{revtex4-1}
\usepackage[T1]{fontenc}
\usepackage[latin9]{inputenc}
\usepackage{bm}
\usepackage{amsmath}
\usepackage{amssymb}

\makeatletter
\usepackage{bbold}

\makeatother

\usepackage{babel}
\begin{document}

\title{Chern-Simons Theory and Dynamics of Composite Fermions}

\author{Junren Shi}
\email{junrenshi@pku.edu.cn}

\selectlanguage{english}%

\affiliation{International Center for Quantum Materials, Peking University, Beijing
100871, China}

\affiliation{Collaborative Innovation Center of Quantum Matter, Beijing 100871,
China}
\begin{abstract}
We propose a $(4+1)$ dimensional Chern-Simons field theoretical description
of the fractional quantum Hall effect. It suggests that composite
fermions reside on a momentum manifold with a nonzero Chern number.
Based on derivations from microscopic wave functions, we further show
that the momentum manifold has a uniformly distributed Berry curvature.
As a result, composite fermions do not follow the ordinary Newtonian
dynamics as commonly believed, but the more general symplectic one.
For a Landau level with the particle-hole symmetry, the theory correctly
predicts its Hall conductance at half-filling as well as the symmetry
between an electron filling fraction and its hole counterpart. 
\end{abstract}
\maketitle
The theory of composite fermions presents one of the most important
insights of the modern condensed matter physics: a collection of strongly
correlated electrons in a partially filled Landau level could be mapped
to a collection of weakly interacting composite fermions (CFs)~\cite{Jain2007}.
While the theory has tremendous successes in understanding the fractional
quantum Hall effect (FQHE) and related phenomena, the true nature
of the CFs is yet to be fully clarified. Conventional wisdom had long
held that a CF behaves just like an ordinary Newtonian particle~\cite{Kalmeyer1992,Halperin1993,Simons1998}.
However, the interpretation suffers from two salient issues: it cannot
correctly predict the Hall conductance of a half-filled Landau level
with the particle-hole symmetry~\cite{Kivelson1997}, and the CF
pictures for an electron filling fraction and its hole counterpart
are seemly asymmetric~\cite{Jain2007}. These issues are recently
addressed by Son, who argues that a CF should be a Dirac particle~\cite{Son2015,Son2016}.
The new interpretation is appealing in many aspects and attracts much
attention, although its microscopic underpinning is not yet clear~\cite{Metlitski2016,Potter2016}.

In this paper, we have a resurvey on the Chern-Simons (CS) theory
of CFs. We argue that a proper CS field theoretical description of
the FQHE should be formulated in $(4+1)$ dimensions, where $4$ stands
for the dimensions of the phase space of CFs. We find that the theory
can describe the FQHE only when the CF momentum manifold within the
phase space has a nonzero Chern number. With a properly assigned Chern
number, the theory correctly predicts the Hall conductance of a half-filled
Landau level with the particle-hole symmetry. Based on derivations
from microscopic wave functions, we further show that the momentum
manifold has a uniformly distributed Berry curvature. As a result,
CFs do not follow the ordinary Newtonian dynamics as commonly believed,
but the more general symplectic one~\cite{Sundaram1999,Xiao2010}.
We explicitly construct a CF effective hamiltonian, which satisfies
all constraints and shows the symmetry between an electron filling
fraction and its hole counterpart. It suggests that the apparent asymmetry
is actually a manifestation of the nontrivial topological structure
of the CF momentum manifold.

\smallskip{}

\noindent \emph{Effective Field Theory.} The deficiency of the conventional
CS theory of the FQHE is actually apparent from the full form of its
Lagrangian which explicitly shows the dependence on the direction
of the external magnetic field $B$~\cite{Zhang1992}:
\begin{equation}
L_{cs}^{(1+2)}=\frac{\mathrm{sign}(B)}{8\pi}\epsilon^{\mu\nu\lambda}a_{\mu}\partial_{\nu}a_{\lambda}-(A+a)_{\mu}j_{cf}^{\mu}+L_{cf},\label{eq:Lcs12}
\end{equation}
where $A^{\mu}\equiv(\phi(\bm{x},t),A_{x_{1}}(\bm{x},t),A_{x_{2}}(\bm{x},t))$
denotes the external electromagnetic field in $(2+1)$ dimensions
with a scalar potential $\phi(\bm{x})$ and a vector potential $\bm{A}(\bm{x})$~\cite{Note1},
$a$ denotes the emergent CS field due to the flux attachment of CFs,
$j_{cf}$ and $L_{cf}$ are CF current density and Lagrangian, respectively,
and we adopt the natural units $q=\hbar=1$, where $q$ is the charge
of a carrier. We note the singular dependence of the coefficient of
the CS term on $B$, and interpret it as an effect of hidden degrees
of freedom.

To remedy the issue, we propose that a proper CS theory for the FQHE
should be formulated in $(4+1)$ dimensions, in the four dimensional
CF phase space that consists of a two-dimensional real space and a
two-dimensional momentum manifold. The effective field theory is described
by the action:
\begin{align}
S & =\int dtd^{2}x\frac{d^{2}k}{(2\pi)^{2}}L_{cs},\label{eq:Action}\\
L_{cs} & =-\frac{1}{8}\epsilon^{\mu\nu\lambda\alpha\beta}a_{\mu}\partial_{\nu}a_{\lambda}\partial_{\alpha}a_{\beta}-(A+a)_{\mu}j_{cf}^{\mu}+L_{cf},\label{eq:Lcs}
\end{align}
where $A$ and $a$ are extended to $(4+1)$ dimensions by:
\begin{align}
A^{\mu} & =(\phi(\bm{x},t),A_{x_{1}}(\bm{x},t),A_{x_{2}}(\bm{x},t),0,0),\\
a^{\mu} & =(a_{0}(\bm{x},t),a_{x_{1}}(\bm{x},t),a_{x_{2}}(\bm{x},t),a_{k_{1}}(\bm{k}),a_{k_{2}}(\bm{k})),\label{eq:acs}
\end{align}
with new Berry connection components, which give rise to a Berry curvature~\cite{Xiao2010}:
\begin{equation}
\Omega(\bm{k})=\partial_{k_{1}}a_{k_{2}}-\partial_{k_{2}}a_{k_{1}},
\end{equation}
and a quantized Chern number:
\begin{equation}
C=\frac{1}{2\pi}\int d^{2}k\Omega(\bm{k}),
\end{equation}
where the integral is over the whole momentum manifold. We require
that the Chern number is related to the external magnetic field by
the constraint:
\begin{equation}
C=\mathrm{sign}(B).\label{eq:Chern}
\end{equation}

From the theory, the coefficient in the conventional CS theory is
identified to be the Chern number $C$ of the CF momentum manifold.
This can be seen by completing the integral over the momentums in
Eq.~(\ref{eq:Action}) for the CS term of Eq.~(\ref{eq:Lcs}). It
recovers the conventional CS form of Eq.~(\ref{eq:Lcs12}) with $C$
as its coefficient. The connection has an important implication: a
CF must reside on a momentum manifold with a nonzero Chern number.
It presents a picture different from the conventional Halperin-Lee-Read
(HLR) theory of CF liquids~\cite{Halperin1993,Simons1998}, for which
the CF momentum manifold is assumed to be identical to that for an
ordinary Newtonian particle in a free space.

The most important consequence of the theory is its correction to
the dynamics of CFs. This is because when the Berry connection is
introduced into the theory, it is inevitable to induce a coupling
between the connection and the momentum components of the current
density, as evident in Eq.~(\ref{eq:Lcs}). It results in a CF dynamics~\cite{Sundaram1999,Xiao2010}:
\begin{align}
\dot{\bm{x}} & =\frac{\partial h_{cf}}{\partial\bm{k}}-\dot{\bm{k}}\times\bm{\Omega}(\bm{k}),\label{eq:xdot}\\
\dot{\bm{k}} & =\bm{E}_{cf}+\dot{\bm{x}}\times\bm{B}_{cf},\label{eq:kdot}
\end{align}
where $\bm{E}_{cf}$, $\bm{B}_{cf}$ are the effective CF electric
and magnetic fields corresponding to $A+a$, respectively, $h_{cf}$
denotes the hamiltonian of CFs, and $\bm{\Omega}(\bm{k})\equiv\Omega(\bm{k})\hat{z}$.
An anomalous velocity correction induced by the Berry curvature is
evident. The presence of the Berry curvature is not surprising because
electrons, which are constituents of CFs, are limited in a Landau
level. It is known that a projection to a truncated Hilbert space
(\emph{e.g.}, a Landau level) is the origin of Berry curvature corrections~\cite{Xiao2010}. 

With the anomalous velocity correction, a CF Fermi sea will have a
finite Hall conductance due to the anomalous Hall effect~\cite{Jungwirth2002}.
This is different from the HLR theory, which interprets the Fermi
sea as a collection of free-electron-like CFs with a zero Hall conductance~\cite{Halperin1993,Simons1998}.
Here, we require that the CF momentum manifold inherits essential
characteristics from its hosting Landau level, including a total phase
volume ($2\pi|B|$), the particle-hole symmetry (in case it has),
as well as the finite Chern number (albeit with an opposite sign~\cite{Note2}).
For a half-filled Landau level with the particle-hole symmetry, the
corresponding CF momentum manifold will also be half-filled and particle-hole
symmetric. In this case, the CF Hall conductance is~\cite{Jungwirth2002,Haldane2004}:
\begin{equation}
\sigma_{xy}^{cf}=-\frac{C}{4\pi},\label{eq:sigmacf}
\end{equation}
where the CF Hall conductance is fixed at a value of one half of the
quantized Hall conductance of a fully occupied momentum manifold,
as a result of the particle-hole symmetry. The topological protection,
assisted by the particle-hole symmetry, makes the value of the CF
Hall conductance robust against perturbations of all possible origins
with the required symmetry. It is easy to verify that the CF Hall
conductance is exactly the one needed for correctly predicting the
Hall conductance of a half-filled Landau level with the particle-hole
symmetry~\cite{Kivelson1997,Son2015}.

\smallskip{}

\noindent \emph{Berry Curvature Distribution. }With the existence
of the Berry curvature established, we proceed to show that the Berry
curvature is actually uniformly distributed in the momentum manifold
with:
\begin{equation}
\Omega(\bm{k})=\frac{1}{B}.\label{eq:BerryCurvature}
\end{equation}
We corroborate the conclusion as follows.

Firstly, we have derived the effective dynamics of CFs in a CF Wigner
crystal directly from its microscopic wave function in Ref.~\cite{Shi2016}.
The dynamics has the exactly same form as Eqs.~(\ref{eq:xdot},\ref{eq:kdot}),
with a Berry curvature that has a value $1/B$ at the limit of $\bm{k}\rightarrow0$~\cite{Note3},
consistent with Eq.~(\ref{eq:BerryCurvature}). The derivation also
shows that the dynamics has the particular form only when the position
of a CF is assigned to the position of its constituent quantum vortices.
Had we assigned the position to its constituent electron, the Berry
curvature would change sign, and other components of Berry curvatures~\cite{Sundaram1999,Xiao2010}
would emerge as long as the emergent CS field does not exactly cancel
the external field. In this case, the effective theory would not be
described by the simplistic form of the CS field Eq.~(\ref{eq:acs})
we have assumed. Hence, to keep our theory simple, it is preferable
to interpret the position of a CF as the position of its constituent
quantum vortices.

Secondly, the conclusion is also consistent with a heuristic argument
based on the dipole picture of CFs~\cite{Simons1998,Read1994,Pasquier1998}.
In the picture, a CF is interpreted as a dipole binding an electron
and two (or an even number of) quantum vortices, with a displacement
$\hat{z}\times\bm{k}/B$ between them. The electron is only coupled
to the external fields, while the quantum vortices are only coupled
to the emergent CS fields~\cite{Metlitski2016,Potter2016,Shi2016}.
Moving a CF along a closed path enclosing an area $S_{\bm{k}}$ in
the momentum manifold is equivalent to fixing the position of its
constituent quantum vortices and moving its constituent electron along
a real space path enclosing an area $S_{\bm{k}}/B^{2}$. Since the
electron is coupled to the external magnetic field, it acquires an
Aharonov-Bohm phase $B\times S_{\bm{k}}/B^{2}=S_{\bm{k}}/B$, which
is exactly the Berry phase expected from the Berry curvature Eq.~(\ref{eq:BerryCurvature}).

Thirdly, we can determine the Berry curvature directly from the Rezayi-Read
wave function~\cite{Rezayi1994}, which is believed to be the correct
microscopic wave function for CF Fermi liquids at half filling. The
wave function has the form:
\begin{equation}
\Psi_{CFS}=\hat{P}_{LLL}e^{-\frac{|B|}{4}\sum_{i}|z_{i}|^{2}}\prod_{i<j}(z_{j}-z_{j})^{2}\Psi_{F}\left(\left\{ \bm{x}_{i}\right\} \right),
\end{equation}
where $z_{i}\equiv x_{1i}+i\mathrm{sign}(B)x_{2i}$, $\Psi_{F}$ denotes
a Slater determinant of plane-wave states of a filled Fermi disc,
and $\hat{P}_{LLL}$ denotes a projection to the lowest Landau level.
To trace the motion of a CF and determine its dynamics, we add one
electron into the Fermi sea and ignore its exchange symmetry with
other electrons. The resulting wave function can be written as:
\begin{equation}
\Psi_{\bm{k}}\left(\bm{x},\left\{ \bm{x}_{i}\right\} \right)\propto e^{i\bm{k}\cdot\bm{x}/2}\Psi_{a}\left(z+ik/B,\left\{ z_{i}\right\} \right),\label{eq:Psik}
\end{equation}
where $\Psi_{a}(z,\{z_{i}\})\propto\exp(-|B||z|^{2}/4)\prod_{i}(z-z_{i})^{2}\Psi_{CFS}$,
$\bm{x}$ and $\bm{k}$ denote the position and momentum of the added
electron, respectively, and $k\equiv k_{1}+i\mathrm{sign}(B)k_{2}$.
One can show that $\{\Psi_{\bm{k}}\}$ forms a set of effective plane-wave
bases with $\left\langle \Psi_{\bm{k}}\left|\Psi_{\bm{k}^{\prime}}\right.\right\rangle =\delta_{\bm{k}\bm{k}^{\prime}}$
by applying the identity:
\begin{equation}
\rho_{a}\left(\bm{x},\bm{x}^{\prime}\right)=\frac{1}{S}e^{-\frac{|B|}{4}\left|\bm{x}-\bm{x}^{\prime}\right|^{2}+\frac{iB}{2}(\bm{x}\times\bm{x}^{\prime})\cdot\hat{z}},\label{eq:rho1}
\end{equation}
where $\rho_{a}\left(\bm{x},\bm{x}^{\prime}\right)\equiv\int\prod_{i}d\bm{x}_{i}\Psi_{a}\left(\bm{x},\left\{ \bm{x}_{i}\right\} \right)\Psi_{a}^{\ast}\left(\bm{x}^{\prime},\left\{ \bm{x}_{i}\right\} \right)$
is the one-body density matrix corresponding to $\Psi_{a}$, and $S$
is the total area the the system. Equation (\ref{eq:rho1}) is a result
of the special analytic structure of the wave function Eq.~(\ref{eq:Psik})
as well as its uniform density distribution in the macroscopic limit~\cite{Giuliani2005,MacDonald1988}.

To determine the CF dynamics, we form a wave-packet state using the
bases $\{\Psi_{\bm{k}}\}$ by $\Psi_{wp}=\int d\bm{k}a(\bm{k},t)\Psi_{\bm{k}}$
with $a(\bm{k},t)=\left|a(\bm{k},t)\right|\exp(-i\gamma(\bm{k},t))$,
where $|a(\bm{k},t)|$ is narrowly distributed in the momentum space
centering at $\bm{k}_{c}$ with a width $|\Delta\bm{k}|\ll1/l_{B}$,
and $l_{B}\equiv1/\sqrt{|B|}$ is the magnetic length. The procedure
for determining the wave-packet dynamics is not much different from
that presented in Ref.~\cite{Sundaram1999}, and will not be repeated
here. The crucial step of the derivation is to relate the CF position
$\bm{x}_{cf}\equiv\left\langle \Psi_{wp}\left|\bm{x}-\bm{k}\times\hat{z}/B\right|\Psi_{wp}\right\rangle $
to the parameters of the wave packet. Note that the position of a
CF is assigned to the position of its constituent quantum vortices,
and is related to the electron position by $\bm{x}_{cf}=\bm{x}-\bm{k}\times\hat{z}/B$.
By applying Eq.~(\ref{eq:rho1}), we obtain:
\begin{equation}
\bm{x}_{cf}=\partial_{\bm{k}_{c}}\gamma(\bm{k}_{c},t)-\frac{1}{2B}\bm{k}_{c}\times\hat{z}.
\end{equation}
It corresponds to a Berry connection $\bm{a}(\bm{k})=-\bm{k}\times\hat{z}/2B$,
which gives rise to the Berry curvature Eq.~(\ref{eq:BerryCurvature}).

\smallskip{}

\noindent \emph{A Model of the CF Hamiltonian. }The above discussions
provide important constraints on how a CF effective hamiltonian should
be constructed. First of all, the hamiltonian should be defined on
a momentum manifold with a finite total area $2\pi|B|$. To this end,
we assume that the momentum manifold is a disc with $|\bm{k}|l_{B}\leq1$,
and a CF manifold comprises two bands defined on the disc. The choice
of the momentum manifold is peculiar since it is not in a form known
for real physical systems, \emph{i.e.}, a Brillouin zone defined on
a torus. Since a CF is a fictitious particle defined in a hidden Hilbert
space~\cite{Jain2009}, the choice should not pose an issue. Moreover,
the choice is the simplest one that is isotropic while accommodating
a particle-hole symmetric CF manifold. Secondly, the CF manifold should
have a uniform distribution of the Berry curvature, and yield a Chern
number as requested by Eq.~(\ref{eq:Chern}). Finally, the CF manifold
should be constructed with the particle-hole symmetry. The combination
of these requirements will not uniquely determine the CF hamiltonian,
and a complete determination may require further microscopic details.
Here we show one of the models that meet all the stated requirements
as well as our preference in the interpretation of simplicity.

Our choice of the CF hamiltonian is a four-band model, defined in
terms of $\bm{p}\equiv\bm{k}l_{B}$:
\begin{align}
H_{cf} & =\epsilon_{0}\left[\begin{array}{cc}
\Lambda\left[(\bm{p}\cdot\bm{\sigma})\sigma_{\tau}(\bm{p}\cdot\bm{\sigma})+\beta\right] & \epsilon_{p}\bm{p}\cdot\bm{\sigma}+\frac{\beta}{2}\\
\epsilon_{p}\bm{p}\cdot\bm{\sigma}+\frac{\beta}{2} & -\Lambda\sigma_{\tau}
\end{array}\right],\label{eq:Hcf}
\end{align}
where $\bm{\sigma}\equiv\left(\sigma_{1},\sigma_{2},\sigma_{3}\right)$
with $\sigma_{i}$ being the Pauli matrices, $\sigma_{\tau}\equiv(\tau+\sigma_{3})/2$,
$\tau=\mathrm{sign}(B)$, $\bm{p}\equiv\left(p_{1},p_{2},\epsilon_{p}\right)$,
$\epsilon_{p}\equiv\sqrt{1-p^{2}}$, $\beta\equiv2B_{cf}/|B|$, $\epsilon_{0}\sim e^{2}/\kappa l_{B}$
is a Coulomb energy scale determining the CF energy, $\Lambda\sim\hbar\omega_{c}/\epsilon_{0}\gg1$
with $\hbar\omega_{c}$ being the cyclotron energy. The model yields
two CF bands with dispersions $\epsilon_{cf}(\bm{p})=\pm\epsilon_{0}\epsilon_{p}$
and two vacuum bands with $\epsilon_{v}(\bm{p})\approx\pm\Lambda\epsilon_{0}$.
The two CF bands together form the CF manifold with $(\epsilon_{cf}/\epsilon_{0})^{2}+p^{2}=1$.
Again, the choice of the CF dispersion is arbitrary and is the simplest
one that meets necessary physical requirements. Had one had a different
preference on the CF dispersion, the function $\epsilon_{p}$ in the
off-diagonal blocks should be replaced. 

In the presence of the effective CF magnetic field $B_{cf}$, the
hamiltonian should be modified by the usual Peierls substitution.
Because $\bm{p}\cdot\bm{\sigma}$ does not commute with $\epsilon_{p}$
in this case, we require that $\epsilon_{p}$ is interpreted as its
Taylor series in $p^{2}$, and a term like $(p^{2})^{N}\bm{p}\cdot\bm{\sigma}$
is interpreted as $(1/2^{N})\sum_{k=0}^{N}C_{N}^{k}(p^{2})^{N-k}\bm{p}\cdot\bm{\sigma}(p^{2})^{k}$.
As a result of the operator symmetrization, we have $\epsilon_{p}(p_{1}\pm ip_{2})|n\rangle\rightarrow[\epsilon_{\sqrt{|\beta|n}}\sqrt{|\beta|n}]|n\mp\mathrm{sign}(\beta)\rangle$,
where $|n\rangle$ denotes a state in $n$-th Landau level. The CF
spectrum is determined to be:

\begin{equation}
\epsilon_{cf}=\epsilon_{0}\begin{cases}
\pm\sqrt{1-|\beta|n}, & n=1,2\dots,\\
\pm1, & CB_{cf}<0.
\end{cases}\label{eq:epsilonn}
\end{equation}

It is interesting to observe that the CF spectrum for $\nu<1/2$ ($CB_{cf}>0$)
and $\nu>1/2$ ($CB_{cf}<0$) are identical except that the latter
one has an extra pairs of Landau levels located at the top and the
bottom of the CF bands. This is consistent with the general consideration
that the simultaneous presences of $\Omega(\bm{k})$ and $B_{cf}$
will modify the measure of the phase space by a factor $D=1-B_{cf}\Omega(\bm{k})$~\cite{Xiao2005},
and as a result, the total number of Landau levels accommodated by
the momentum manifold will be reduced or increased by one for the
cases $CB_{cf}>0$ and $CB_{cf}<0$, respectively. Therefore, to make
a filling fraction $\nu_{n}=n/(2n+1)$ symmetric with its hole counterpart
$1-\nu_{n}$, it should be mapped to $n$ CF Landau levels while the
hole counterpart be mapped to $n+1$ levels, to compensate the extraneous
level emerged at the band bottom for the latter case. The consideration
indicates that the apparent asymmetry in the CF pictures for for $\nu<1/2$
and $\nu>1/2$ is not a flaw of the CF theory. Instead, it is a manifestation
of the non-zero Chern number of the CF manifold and a necessary feature
to make the physics symmetric in such a manifold.

Finally, the model can be transformed by $\tau\rightarrow-\tau$.
The resulting model has a bottom band with a nonzero vacuum Chern
number $C_{v}=\mathrm{sign}(B)$, as well as a CF manifold with the
Chern number $C=-\mathrm{sign}(B)$. It is also possible to construct
a valid effective theory based on the model. In this case, the Lagrangian
of the effective theory should be written as:
\begin{equation}
L_{cs}^{(v)}=\bar{L}_{cs}-\frac{C_{v}}{2|B|}\epsilon^{\mu\nu\lambda}(A+a)_{\mu}\partial_{\nu}(A+a)_{\lambda},\label{eq:Lcsv-1}
\end{equation}
where the extra term accounts for the contribution of the non-trivial
vacuum, $\bar{L}_{cs}$ denotes the Lagrangian Eq.~(\ref{eq:Lcs})
with the signs of its CS term and the Chern number flipped, and $\mu,\nu,\lambda\in(0,1,2)$.
Actually, Equation~(\ref{eq:Lcs}) and (\ref{eq:Lcsv-1}) correspond
to the views of the FQHE from the electron side and the hole side,
respectively. For the latter, the vacuum consists of all unoccupied
Landau levels which do have a nonzero total Chern number that cancels
the Chern number of the occupied lowest Landau level.

\smallskip{}

\noindent \emph{Relation to the Dirac Theory. }We cannot make our
theory compatible with Son's Dirac theory of CFs. As shown in Ref.~\cite{Son2015},
the Dirac theory can be converted to the conventional HLR theory by
introducing a mass, which opens a gap and effectively introduce a
vacuum with a half-quantized Chern number $C_{v}$. Unfortunately,
the mass will inevitably induce a Berry curvature and a half-quantized
Chern number $-C_{v}$ in the CF manifold. In our theory, unlike the
conventional HLR theory, the effect of the Berry curvature cannot
be eliminated by making the mass infinitely large because the CF momentum
manifold is finite and particle-hole symmetric, and the Berry curvature
has nowhere to disappear. To obtain the correct CF Hall conductance
(zero in this case), one has to assume that the Chern number is exactly
cancelled, resulting in a CF manifold with a zero Chern number. This
is incompatible with the effective theory Eq.~(\ref{eq:Lcs}), for
which a zero Chern number cannot provide a valid description for the
FQHE.

The most important difference between our theory and the Dirac theory
is in how the correct CF Hall conductance at the half filling is obtained
and protected. In our theory, it is induced by the CF anomalous Hall
effect and protected by the particle-hole symmetry of the CF manifold.
On the other hand, in the Dirac theory, the half quantized Hall conductance
is originated and topologically protected at the Dirac point. However,
the CF Fermi level corresponding to the half-filling is far removed
from the Dirac point. It seems to be nontrivial to find a mechanism
to prevent a non-vanishing Hall conductance from arising out of the
Fermi sea of CFs in all possible scenarios, for instance, when the
full effect of disorders as discussed in Ref.~\cite{Wang2016} is
considered.

In summary, we have constructed a new form of CS theory for CFs in
$(4+1)$ dimensions and determined the topological structure of the
CF momentum manifold. It resolves all the salient issues which had
been interpreted as the flaws of the CF theory. Our theory suggests
that they are actually the manifestations of the non-trivial topological
structure of the CF manifold and the necessary features of a proper
CF theory. 
\begin{acknowledgments}
This work is supported by National Basic Research Program of China
(973 Program) Grant No. 2015CB921101 and National Science Foundation
of China Grant No. 11325416.
\end{acknowledgments}

\end{document}